\documentclass[reprint,superscriptaddress,prb,aps]{revtex4-2}

\usepackage{mathtools}
\usepackage{amsmath}
\usepackage{amsthm}
\usepackage{amssymb}
\usepackage{graphicx}
\usepackage{bm}
\usepackage[colorlinks,linkcolor=blue, urlcolor=blue,citecolor=blue,bookmarks=false]{hyperref}

\begin{document}

\title{Behavior under magnetic field of resonance \\ at the edge of the upper Hubbard band in 1$T$-TaS$_{2}$}

\author{C. J. Butler}
\email{christopher.butler@riken.jp}
\affiliation{RIKEN Center for Emergent Matter Science, 2-1 Hirosawa, Wako, Saitama 351-0198, Japan}

\author{M. Yoshida}
\affiliation{RIKEN Center for Emergent Matter Science, 2-1 Hirosawa, Wako, Saitama 351-0198, Japan}

\author{T. Hanaguri}
\email{hanaguri@riken.jp}
\affiliation{RIKEN Center for Emergent Matter Science, 2-1 Hirosawa, Wako, Saitama 351-0198, Japan}

\author{Y. Iwasa}
\affiliation{RIKEN Center for Emergent Matter Science, 2-1 Hirosawa, Wako, Saitama 351-0198, Japan}
\affiliation{Quantum-Phase Electronics Center and Department of Applied Physics, The University of Tokyo, 7-3-1 Hongo, Bunkyo-ku, Tokyo 113-8656, Japan}

\begin{abstract}
Recent theoretical investigations of quantum spin liquids have described phenomenology amenable to experimental observation using scanning tunneling microscopy. This includes characteristic resonances found at the edge of the upper Hubbard band of the host Mott insulator, that under certain conditions shift into the Mott gap under external magnetic field [W.-Y. He and P. A. Lee, \href{https://doi.org/10.48550/arXiv.2212.08767}{arXiv:2212.08767}]. In light of this we report scanning tunneling microscopy observations, in samples of the quantum spin liquid candidate 1$T$-TaS$_{2}$, of a conductance peak at the upper Hubbard band edge and its magnetic field dependent behavior. These observations potentially represent evidence for the existence of a quantum spin liquid in 1$T$-TaS$_{2}$. We also observe samples in which such field dependence is absent, but with no observed correlate for the presence or absence of field dependence. This suggests one or more material properties controlling electronic behavior that are yet to be understood, and should help to  motivate renewed investigation of the microscopic degrees of freedom in play in 1$T$-TaS$_{2}$, as well as the possible realization of a quantum spin liquid phase.
\end{abstract}

\maketitle

\section{Introduction}

The transition metal dichalcogenide 1$T$-TaS$_{2}$ is one among several candidate materials to host a quantum spin liquid (QSL) phase. At low temperature it features a 3Q commensurate charge density wave (CDW), equivalently described as a triangular lattice of `star-of-David' polaronic clusters \cite{Wilson1975, Fazekas1979, Fazekas1980}. Each cluster contains an odd number of orbitals which, if interlayer interactions are neglected, leaves the system at half-filling with one leftover spin per cluster. This suggests that the material's insulating behavior is due to Mott localization, and given this, the resulting array of unpaired spins of \textonehalf\ localized and geometrically frustrated on a triangular lattice provides a possible setting for a QSL ground state \cite{Law2017}. The absence of conventional magnetic order in the bulk down to very low temperature, consistent with the presence of a spin liquid, has been confirmed using $\mu$-SR measurements \cite{Klanjsek2017}, and very recent spin-polarized STM measurements have also indicated spin frustration at the surface \cite{Lee2023}.

Collecting affirmative evidence for a QSL ground state is challenging because it has no associated local order parameter, and is instead most firmly characterized by non-local properties such as the entanglement entropy \cite{Anderson1973, Zhou2017, Jiang2012}. However, recent theoretical explorations have suggested that the characteristic fractional excitations of a QSL, such as spinons, can be accompanied by charge modulations, bound states, or inelastic tunneling events that are amenable to detection using a local electronic probe such as a scanning tunneling microscope (STM) \cite{Tang2013, Ruan2021, Chen2022, He2022a, He2022b, Konig2020, Udagawa2021}.
Some of these phenomena have been reported for 1$T$-TaSe$_{2}$. In monolayer 1$T$-TaSe$_{2}$ samples, charge modulations predicted to accompany an instability of the spinon Fermi surface, and Kondo resonance consistent with itinerant spinons coupling to magnetic adatoms, have both been reported \cite{Ruan2021, Chen2022}.

Further theoretical predictions of locally observable phenomena characteristic of a QSL have very recently been put forward by He and Lee \cite{He2022a, He2022b}. In particular, they describe a resonance at the lower edge of the upper Hubbard band (UHB) of the host Mott insulator, which results from the propensity of fluctuations of the emergent gauge field to bind spinons and chargons into electrons (`gauge binding') \cite{He2022a}. Furthermore, they calculate the effect of Landau quantization of the spinon Fermi surface, and show that the resonance can shift with the application of magnetic field. Experimental observations of this behavior may serve as a signifier of QSL behavior.

In this work we report on a narrow peak in tunneling conductance at the UHB edge in 1$T$-TaS$_{2}$, and its magnetic field dependence, exhibiting the behavior described by the theory put forward by He and Lee. We establish a simple quantitative characterization of the observations, estimating power law parameters that describe the field dependent energy shift of the peak. We also show that such magnetic field dependence is not always present, suggesting that it is contingent upon one or more degrees of freedom that are not readily observable in our STM measurements. With this in mind, and briefly reviewing recent reports on interlayer stacking and its effects on surface local density-of-states, we discuss the need for greater understanding of these interlayer effects and the ramifications for the properties of any existing QSL in the bulk or at the surface of 1$T$-TaS$_{2}$.

\section{Results}

Crystals of 1$T$-TaS$_{2}$ were synthesized as described previously \cite{Tani1981}, and prepared for STM measurements by cleaving at $\approx$77~K in ultra-high vacuum ($P \approx 10^{-10}$ Torr), before insertion into a modified Unisoku USM1300 low-temperature STM system held at $T$ = 1.5~K \cite{Hanaguri2006}. STM measurements were performed using electro-chemically etched tungsten tips, which were characterized and conditioned using field ion microscopy followed by mild indentation at a clean Cu(111) surface. Tunneling conductance was measured using the lock-in technique with bias modulation of frequency $f_{\textrm{mod}}$ = 617.3~Hz and amplitude $V_{\textrm{mod}}$ = 1~mV.

\begin{figure}
\centering
\includegraphics[scale=1]{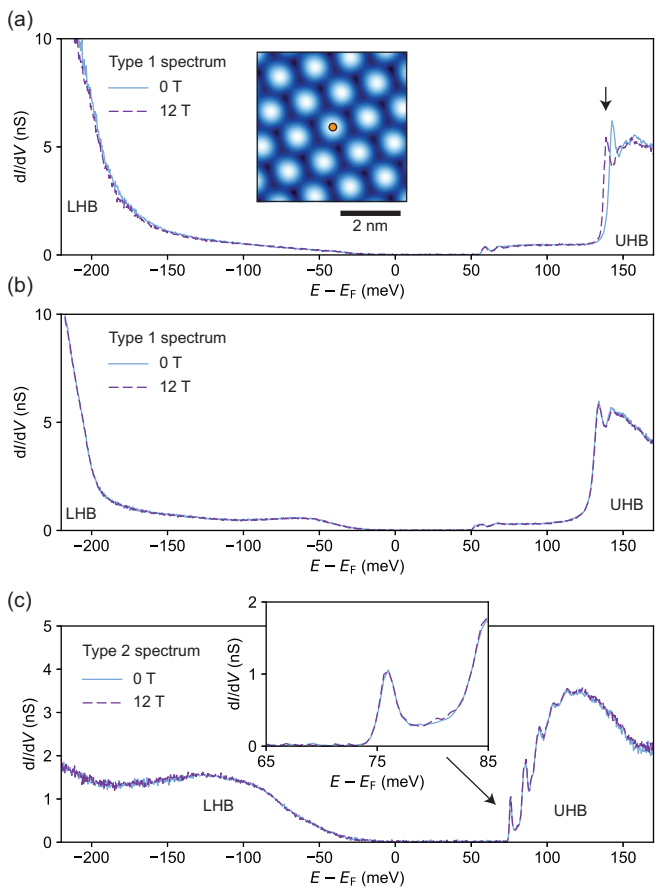}
\caption{\label{fig:1}
\textbf{Tunneling conductance curves under zero and high magnetic field.} (a) Conductance curves acquired atop one star-of-David cluster at a type 1 surface found on sample S2 (at the location shown in the inset, setpoints $V$ = -500~mV, $I$ = 10~pA). Data measured at $\mu_{\mathrm{B}}H$ = 0~T and 12~T are shown. The spectroscopic feature under consideration is marked by a black arrow. (b) Conductance curves acquired in the same way at another type 1 surface on sample S1, here showing an absence of magnetic field dependence in the corresponding spectroscopic feature. (c) Conductance curve acquired in the same way at a type 2 surface. The inset shows an expanded view of the lowest-energy peak in the UHB, showing absence of field dependence. (The peaks at higher energies have been attributed to electron-phonon sidebands \cite{Butler2021}.)
}
\end{figure}

The relevance of interlayer interactions, mediated by interlayer stacking degrees of freedom, for the electronic structure of 1$T$-TaS$_{2}$ and for the presence and the nature of any extant QSL phas has recently gained prominence \cite{Ritschel2015, Cho2016, Ma2016, Ritschel2018, Lee2019, Butler2020, Wu2022, ManasValero2021}. An overview of the current understanding of these interlayer effects puts the surface density-of-states spectra measured using STM in the proper context. Figures 1(a) and (b) show tunneling spectra belonging to one of the two broad categories that have been observed at the 1$T$-TaS$_{2}$ surface, while Figure 1(c) shows a spectrum belonging to the other category. The existence of distinct categories is argued to result from interlayer stacking effects in bulk or multilayer 1$T$-TaS$_{2}$. The bulk CDW stacking configuration is likely some version of a staggered bilayer stacking pattern \cite{LeGuyader2017, vonWitte2019, Stahl2020, Butler2020}. Taken at face value this leads to the possibility that bulk 1$T$-TaS$_{2}$ might be a dimerized Mott insulator or even a band insulator \cite{Ritschel2018, Lee2019}. A correspondence between the surface density-of-states and the stacking of the surface layer atop the first buried layer was previously suggested \cite{Butler2020}, identifying `type 1' [Figs. 1(a) and (b)] and `type 2' [Fig. 1(c)] spectra thought to be characteristic of the dimer-preserving and dimer-breaking surface terminations of the bilayer stacking pattern. But a subsequent report by Wu \textit{et al.} has hinted that knowledge of the near-surface stacking is insufficient to determine the surface density of states \cite{Wu2022}, possibly implying that there are relevant `hidden' degrees of freedom that cannot readily be characterized with the methods used so far.

As an aside, Wu \textit{et al.} also show that the commonly observed `large gap' conductance spectrum (type 1) appears to show a degree of resilience against the effects of interlayer interactions, and therefore suggest that it represents the density of states intrinsic to a single layer \cite{Wu2022}. While this is contradictory to some previously mentioned results \cite{Lee2021, Petocchi2022, Jung2022}, it does reconcile with the large gap reported by V\v{a}no \textit{et al.} for monolayer 1$T$-TaS$_{2}$ \cite{Vano2021}.

Each series of conductance curves presented in this work was acquired atop a single star-of-David cluster chosen in the given location, as shown in the inset to Fig. 1(a). Effort was made to perform all series of measurements in defect-free regions, and we therefore assume that the curves presented here represent the typical and intrinsic behavior of clusters within the given domain. For each of the tunneling spectra shown in Figure 1, the UHB exhibits a narrow peak at its lower edge, marked by a black arrow in Fig. 1(a). Previous work has demonstrated the ubiquity of this feature in defect-free regions, visualizing the its spatial distribution over fields-of-view encompassing multiple clusters, and within individual clusters \cite{Butler2021}. In each case we show tunneling conductance curves acquired first under zero magnetic field and again under a field of $\mu_{\mathrm{B}}H$ = 12~T applied perpendicular to the sample surface.

In the case of Fig. 1(a) (sample S2), this feature undergoes a shift toward lower energy at $\mu_{\mathrm{B}}H$ = 12~T. Observation of a similar finite energy shift of the peak at the UHB edge was reproduced for two samples (S2 and S3), and for two CDW domains on S3 (D1 and D2) for a total of three instances. However Fig. 1(b) shows that in another sample (S1), a magnetic field dependent energy shift of the peak was seen to be absent. This absence of the field-induced shift was observed in two type 1 domains found at one cleaved surface.
Figure 1(c) shows results of a similar measurement at the previously described `type 2' surface, which also shows an absence of magnetic field dependence. For the type 2 spectrum, shown in Fig. 1(c) additional peaks at higher energies are thought to be replicas due to electron-phonon coupling involving the amplitude mode of the CDW \cite{Butler2021}. The possible origin of the lowest-lying peak, and for the peak in the type 1 spectrum, is described in the Discussion section below. 

In Figs. 1(a) and 1(b), a subtle spectroscopic feature appears between $\sim$50~meV and the main UHB. This feature does not show any field dependence and was not observed to have a correlation with the presence or absence of field dependence in the peak at the UHB edge. This feature is not attributable to the presence of a point defect, and may instead be related to near-surface inter-layer stacking effects that are not amenable to characterization with the methods used here.

Below we focus on the detailed behavior of the UHB peak in those samples that displayed significant magnetic field dependence. Figure 2 summarizes the numerical fitting procedures for estimating the energy shift of the peak and, for comparison, the LHB onset. The raw conductance curves collected on sample S2, both at the LHB onset and at the UHB edge, between $\mu_{\mathrm{B}}H$ = 0~T and 12~T in increments of 1~T, are shown in Figs. 2(a) and 2(b). In order to track the magnetic field dependence of the spectra, for each value of magnetic field we use fitting to extract both the energy of the peak at the UHB edge and a value of energy for the LHB onset. 

For the onset of the LHB we use a linear fit to a segment of the slope between -210 and -190~meV. Using the gradient $\frac{\mathrm{d}g}{\mathrm{d}E}$, we convert the average conductance within this range, which we name $\tilde{g}$, to an energy \textit{via} $ \tilde{E} = - \frac{\mathrm{d}E}{\mathrm{d}g} \tilde{g} $. The absolute value of $\tilde{E}$ is irrelevant because we will consider only the variation with respect to the zero field value, $\Delta E_{\mathrm{LHB}}(H) = \tilde{E}(H) - \tilde{E}(H=0)$.

We model the conductance spectrum for the UHB edge as a Lorentzian lineshape, representing the narrow peak, on a hypertangent function that represents the onset of the UHB continuum:

\begin{align}
    g(E) = &\frac{a}{\pi} \frac{(\Gamma / 2)}{(E - E_{\mathrm{peak}})^{2} + (\Gamma / 2)^{2}} \\ 
    &+ b(1 + \tanh(w(E - E_{\mathrm{edge}}))) \notag\\
    &+ c . \notag
\end{align}

\begin{figure}
\centering
\includegraphics[scale=1]{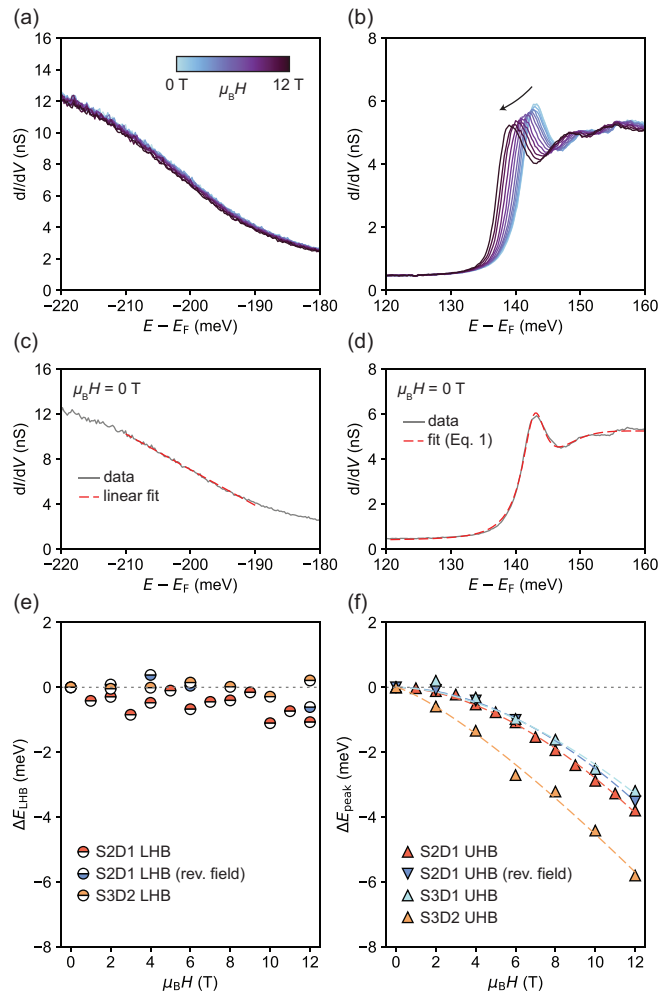}
\caption{\label{fig:2}
\textbf{Estimation of field dependent energy shifts of the UHB peak.} (a) and (b) Expanded views of conductance curves acquired at the LHB onset and the UHB edge, respectively, for S2 [see Fig. 1(a)]. The magnetic field was varied from 0 to 12~T, in increments of 1~T. (c) A linear fit to the LHB onset as described in the main text, and (d) fitting of the function described by Eq. 1 to the data at the UHB edge, obtaining the estimated $E_{\mathrm{peak}}$, in this case for the zero field data. (e) and (f) Energy shifts for both the UHB peak and LHB features, $\Delta E_{\mathrm{peak}}$ and $\Delta E_{\mathrm{LHB}}$, relative to the zero field data. Data are included for two samples (S2 and S3), and for two CDW domains on S3 (D1 and D2). The dark blue inverted triangles and dark blue half-filled circles correspond to a measurement with reversed magnetic field with respect to the others. The dashed lines represent results of a power law fit (Eq. 2) to each series of points for the UHB peak.}
\end{figure}

\begin{table}
\caption{\label{tab:table1}%
Power law parameters describing $\Delta E_{\mathrm{peak}}(H)$.}
\begin{ruledtabular}
\begin{tabular}{lll}
\textrm{Data series} &
\textrm{$\alpha$} &
\textrm{$\beta$} \\
\colrule
S2D1 & -0.047 & 1.77 \\
S2D1 (rev. field) & -0.033 & 1.88 \\
S3D1 & -0.041 & 1.76 \\
S3D2 & -0.249 & 1.26 \\
\end{tabular}
\end{ruledtabular}
\end{table}

The functions fitted to the LHB and UHB data at zero field are shown in Figs. 2(c) and 2(d). The relevant energies of the respective LHB and UHB features extracted from these procedures, for all measured magnetic field values and all relevant samples, are shown in Figs. 2(e) and 2(f). Also included is one series, acquired on S2, for which the magnetic field polarity was reversed in comparison to the others (dark blue markers). In each case we plot the energy variation with respect to the energy of the relevant feature in the zero field data, i.e. $\Delta E_{\mathrm{peak}}(H) \equiv E_{\mathrm{peak}}(H) - E_{\mathrm{peak}}(H=0)$ for the UHB peak, and also the shift of the LHB as described above. Fig. 2(e) shows at most a very small magnetic field dependent shift of the LHB as a whole.

For the field dependence of the UHB peak, to quantitatively characterize the observed phenomena, we also fit a simple power law,

\begin{equation}
\Delta E_{\mathrm{peak}}(H) = \alpha(\mu_{\mathrm{B}}H)^{\beta} ,
\end{equation}

through each series of points. Here $\mu_{\mathrm{B}}$ is the Bohr magneton. The fitting results are collated in Table 1.

\section{Discussion}

Aspects of the observed magnetic field dependence of the conductance peak at the UHB edge appear to disfavor prosaic explanations. The observation of a shift rather than a splitting appears to exclude the Zeeman effect. A shift could occur for a fully spin-polarized state, but it is very unlikely that such a state exists in the UHB of 1$T$-TaS$_{2}$ and, as mentioned above, the possibility of a nearby magnetic impurity (or any other type of impurity) was carefully excluded. The nonlinear ($1 < \beta < 2$) relation of the peak energy with $H$ also argues against the Zeeman effect as an explanation. Also, the direction of the shift ($\alpha < 0$) is opposite to that expected for the behavior of a Landau level formed at the bottom of an electron band.

The observation of a peak at the UHB edge is consistent with the predicted resonance described in Ref.\cite{He2022a}, resulting from quantization of a spinon band followed by weak attraction of spinons and chargons that binds them into electrons (gauge binding). Its appearance only at the UHB edge but not at the LHB edge would, in this picture, be related to the details of the dispersion of the underlying spinon band.

The observed magnetic field dependence of the peak, displayed in Fig. 2, showing decreasing energy with increasing magnetic field, is consistent with behavior predicted in the distinct regime of strong gauge binding also described in Ref.\cite{He2022a}. An intuitive description of why the peak energy decreases is that the magnetic field acts to quantize the spinon band, also causing spinon wavefunctions to become increasingly localized. Given an attractive interaction between spinons and chargons, the binding of a chargon (specifically a doublon, for the UHB) with a more spatially localized spinon becomes increasingly energetically favorable with increasing magnetic field.

This field dependence has been reproduced at multiple measured surfaces, but interestingly is absent for other surfaces. Adopting the model put forward in Ref.\cite{He2022a} in order to interpret the present observations, it must be emphasized that it is the presence of the narrow resonance at the UHB that signifies the existence of a QSL, regardless of the subsequent observation of magnetic field dependence. The magnetic field dependence, however, is the feature that sets the above mentioned model apart from other potential explanations for a peak at the UHB edge. To the best of our knowledge no other potential explanations for the peak, such as hypothesized heavy quasiparticles \cite{Granath2014} or possible polaronic bound states, accommodate magnetic field effects such as those observed.

Within the framework of Ref.\cite{He2022a}, the detailed magnetic field dependent peak energy would indicate a regime of strong gauge binding ($\beta = 2$), or weak gauge binding ($\beta = 1$). The estimated values of $\beta$ shown in Table 1 would correspond to an intermediate situation between strong and weak gauge binding, which the methods in Ref.\cite{He2022a} cannot access, but which generally might manifest in a real material system. 

The current observation indicates that there is a degree of freedom that influences the strength of the binding that has no discernible correspondence to features that can be observed in the top layer using STM. We therefore consider the possibility that interlayer effects play a role. As mentioned above, a correspondence between the surface density-of-states and the stacking of the surface layer atop the first buried layer was previously suggested \cite{Butler2020}, and this was supported by further experimental and numerical results \cite{Lee2021, Petocchi2022, Jung2022}. However, recently Wu \textit{et al.} reported an extensive collection of STM observations showing that a simple 1:1 correspondence between surface electronic structure and near-surface interlayer stacking cannot yet be reached \cite{Wu2022}. Most relevant to the current discussion, surfaces that showed spectra characterized by a larger gap, which we have referred to as type 1 [as in Figs. 1(a) and (b)], were seen to occur for various near-surface stacking configurations. This admits the possibility of one or more `hidden' (i.e. remaining undetermined upon observation of a type 1 spectrum) degrees of freedom that are associated with interlayer stacking, possibly including stacking involving layers beyond the first buried layer. We speculate that the presence or absence of the observed magnetic field dependence might also be controlled by such degrees of freedom, which have not yet been fully characterized by STM experiments. This scenario could also account for the field-insensitive peaks seen at the UHB edge of the type 2 surface. At this stage the relation between the field-dependence and these conjectured degrees of freedom is unclear. Hence there remain avenues for further detailed investigations of interlayer effects and their correlation with such phenomena as the residual density of states within the Mott gap, the peak at the UHB edge, the magnetic field dependence of this peak, as well as others.

In summary, the above finding of field dependence of the UHB edge resonance, taken together with the recent theoretical predictions of He and Lee, may provide evidence for the presence of a QSL in 1$T$-TaS$_{2}$. Given the observation that this field dependence can be absent at some surfaces, and that interlayer effects may introduce complications that remain poorly understood, this should encourage efforts to detect a similar UHB edge resonance in monolayer 1$T$-TaS$_{2}$ and, if it is found, characterize its behavior under external magnetic field.

\section*{Acknowledgements}
We are grateful to Y. Kohsaka, T. Machida, J. Lee, H.-W. Yeom and P. A. Lee for helpful discussions. C.J.B. acknowledges support from RIKEN's Programs for Junior Scientists. This work was supported in part by JSPS KAKENHI grant numbers JP18K13511, JP19H00653, JP19H01855 and JP19H05602.

\section*{Data availability}
The data that support the findings presented here are available from the corresponding authors upon reasonable request.

\end{document}